\documentclass{aa}  

\usepackage{graphicx} % ,newtxmath}	% Including figure files
\usepackage{amsmath,newtxtext}	% Advanced maths commands
\usepackage{txfonts}
\usepackage{latexsym}
\usepackage[dvipsnames]{xcolor}
\usepackage{chngcntr}
\usepackage{xcolor}
\usepackage{soul}

\usepackage{hyperref}
\hypersetup{
    colorlinks=true,
    linkcolor=cyan,
    citecolor=cyan,
    urlcolor=cyan,
    }

\newcommand{\orcid}[1]{\protect\href{https://orcid.org/#1}{\protect\includegraphics[width=8pt]{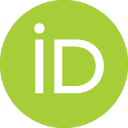}}}

\begin{document}

\title{CGCS\,6306, another X-ray-emitting asymptotic giant branch star confirmed to be a symbiotic binary}
\titlerunning{CGCS\,6306, X-ray-emitting symbiotic binary}

\author{Jaroslav Merc\orcid{0000-0001-6355-2468}
          \inst{1,2}
          \and
          Mart\'\i n A.\ Guerrero\orcid{0000-0002-7759-106X}
          \inst{3}
          \and
          Jes\'us A.\ Toal\'a\orcid{0000-0002-5406-0813}
          \inst{4}
          \and
          Roberto Ortiz\orcid{0000-0002-0084-8373}
          \inst{5}}
\authorrunning{Jaroslav Merc et al.}
   \institute{
Astronomical Institute of Charles University, V Hole\v{s}ovi\v{c}k{\'a}ch 2, Prague, 18000, Czech Republic
\and
Instituto de Astrof\'isica de Canarias, Calle Vía Láctea, s/n, E-38205 La Laguna, Tenerife, Spain\\
\email{jaroslav.merc@mff.cuni.cz}
\and
Instituto de Astrof{\'i}sica de Andaluc{\'i}a, IAA-CSIC, Glorieta de la Astronom\'{i}a S/N, Granada 18008, Spain 
\and
Instituto de Radioastronom\'{i}a y Astrof\'{i}sica, Universidad Nacional Aut\'{o}noma de M\'{e}xico, Morelia 58089, Mich., Mexico
\and
Escola de Artes, Ciências e Humanidades, USP, 
Av.\ Arlindo Bettio 1000, 03828-000 São Paulo, Brazil
}

\date{Received \today; accepted XXX}

% \abstract{}{}{}{}{} 
% 5 {} token are mandatory
% context heading (optional)
% {} leave it empty if necessary  

\abstract{ % Context
A number of asymptotic giant branch (AGB) stars are known to exhibit UV excess and/or X-ray emission.  
These have been considered signposts of a hot white dwarf (WD) companion in a symbiotic system (SySt), but AGB stars are so bright that they easily outshine these companions hampering their detection at optical wavelengths. 
A recent multi-wavelength investigation on the X-ray-emitting AGB (X-AGB) star Y\,Gem has confirmed the presence of a WD companion and, thus, its SySt nature. 
}{ % Aim
Our goal is to explore the true nature of another X-AGB star, namely CGCS\,6306, to investigate whether some objects from this group may in fact be unnoticed symbiotic systems with AGB donors.
}{ % Methods
Optical spectra and photometric data, together with X-ray observations, have been analyzed to investigate the properties of the stellar components and accretion process in CGCS\,6306. 
}{ % Results
CGCS\,6306 is a carbon Mira with a pulsation period of 362 days.  
Its optical spectrum exhibits the typical saw-shaped features of molecular absorptions in addition to H~{\sc i} and He~{\sc i} recombination and [O~{\sc i}] and [O~{\sc iii}] forbidden emission lines. 
The H$\alpha$ line profile is broad, which can be interpreted as evidence for an accretion disk. 
The X-ray spectrum is hard, typical of highly-extincted hot plasma emission, and the X-ray luminosity is $\approx10^{32}$ erg~s$^{-1}$. 
}{ % Conclusions
The detection of high-excitation optical emission lines and the X-ray properties of CGCS\,6306 confirm the presence of a WD companion, making it a bona-fide $\delta$-type X-SySt.  
Its X-ray luminosity is comparable to that of Y\,Gem, the other X-AGB confirmed to be a SySt, which was found to exhibit a high accretion rate.  
The lack of suitable information on the UV and blue optical properties of CGCS\,6306, however, precludes a definitive estimate of the accretion rate in this system. 
Since CGCS\,6306 is a carbon Mira, it adds to the small group of Galactic carbon SySts. 
}
% context heading (optional)
% {} leave it empty if necessary  


\maketitle

%%%%%%%%%%%%%%%%%%%%%%%%%%%%%%%%%%%%%%%%%%%%%%%%%%

%%%%%%%%%%%%%%%%% BODY OF PAPER %%%%%%%%%%%%%%%%%%

\section{Introduction}
Asymptotic giant branch (AGB) stars are the descendants of low- and intermediate-mass stars just before they expel their H-rich envelopes to become planetary nebulae (PNe) and continue their evolution as white dwarfs (WDs).  
While late-type dwarf stars are well known to exhibit coronal-powered X-ray emission \citep{Ayres1981}, AGB stars are not expected to possess coronae because of their large radii and slow rotation.  
Yet X-ray emission has been detected in 47 AGB stars \citep[hereafter X-AGB,][]{Guerrero+2024}.

The detection of X-ray emission in X-AGBs is considered a signpost of binarity, with X-ray emission being produced in an accretion disc around a companion or in the corona of a late-type dwarf secondary \citep{OG2021}.  
The direct detection of the companion in the optical or infrared is, however, hampered by the humongous luminosity of the AGB star, even though far-UV emission or significant near-UV excess would hint to a companion \citep[e.g.,][]{OG2016,Sahai2022}. 
WD companions have been confirmed in a number of symbiotic stars (SySts) with AGB donors \citep[e.g., StHa 32, RX Pup, R Aqr, and others; see the \textit{New Online Database of Symbiotic Variables} and references therein,][]{2019RNAAS...3...28M,2019AN....340..598M}. 
It must be noted, though, that a number of AGB stars are classified as SySt candidates based only on their X-ray emission \citep[see, e.g., the cases of CGCS\,5926, OGLE BLG654.20\,36111, or V371\,CrA;][]{2011A&A...534A..89M,2014ApJ...780...11H,2021MNRAS.506.5619W,2022A&A...661A..38P}.  
These putative X-SySts will thus add to the known sample of X-AGBs where the presence of a compact companion is so far unconfirmed.

The classical diagnostic for the SySt nature of a giant star relies on the detection of optical emission lines from high-ionization species requiring above 35 eV (e.g., He~{\sc ii} $\lambda$4686 \AA, [Fe~{\sc vii}] $\lambda\lambda$5727,6087 \AA, or [O~{\sc iii}] $\lambda$5007 \AA) and/or O~{\sc vi} Raman-scattered lines at 6830 and 7088 \AA\ \citep{Mikolajewska+1997,Belczynski+2000,Miszalski+2013,Merc+2021,Merc2025}. 
We have recently used optical intermediate-dispersion spectroscopic observations to uncover the SySt nature of Y\,Gem \citep{Guerrero+2025}, an AGB star with variable far-UV and very hard and also variable X-ray emissions \citep{Sahai+2008,OG2021} where the presence of a companion was also suggested by flickering of its UV continuum \citep{Sahai+2018}.

\begin{table*}
\centering
\caption{XMM-Newton and Chandra CGCS\,6306 X-ray count rates.}
\label{tbl.x}
\small
\begin{tabular}{lccccc}
\hline
Instrument & Epoch & \multicolumn{4}{c}{Count rate} \\
 & & \multicolumn{4}{c}{(ks$^{-1}$)} \\
 & & \multicolumn{4}{c}{\underline{{~~~~~~~~~~~~~~~~~~~~~~~~~~~~~~~~~~~~~~~Energy Band~~~~~~~~~~~~~~~~~~~~~~~~~~~~~~~~~~~~~~~}}} \\
 & & 0.3--2.0 keV & 2.0--5.0 keV & 5.0--10.0 keV & 0.35--10.0 keV \\
\hline
XMM EPIC MOS1   & 2010.33 & $<0.5$   & $\dots$       & 0.7$\pm$0.6  & 1.2$\pm$0.5\\
XMM EPIC MOS2   & 2010.33 & $\dots$  & $\dots$       & 0.5$\pm$0.6 & 1.4$\pm$1.0 \\
CXO ACIS S3 & 2023.99 & $<0.04$  & 0.31$\pm$0.17 & 0.49$\pm$0.21 & 0.76$\pm$0.26 \\
CXO ACIS S3 & 2024.01 & $<0.005$ & 1.14$\pm$0.27 & 0.48$\pm$0.18 & 1.59$\pm$0.33 \\
\hline
\end{tabular}
\end{table*}

\begin{figure}
\begin{center}
\includegraphics[width=1.0\columnwidth]{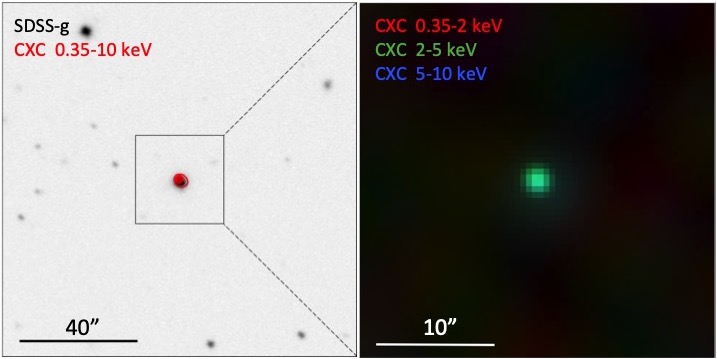}
\caption{
(left) Inverted grayscale $g$-SDSS image of CGCS\,6306 overlaid with CXO ACIS-S X-ray contours in red.  
The field of view is $2^\prime\times2^\prime$. 
(right) Color-composite CXO X-ray picture (R = 0.35-2.0 keV, G = 2.0-5.0 keV, B = 5.0-9.0 keV).  
The field of view is $30^{\prime\prime}\times30^{\prime\prime}$. 
In both images, North is up and East to the left. 
}
\label{img.x}
\end{center}
\end{figure}

The final confirmation of Y\,Gem as a SySt indicates that the acquisition of high-quality optical spectra of X-AGB stars could be a new pathway to unveil compact companions wherever the giant star overshines it. 
To further strengthen this premise before starting a long-term project to acquire optical spectroscopy of X-AGB stars, 
we present here a second case-study of a previously unexamined X-AGB star, namely CGCS\,6306 \citep[aka IRAS\,08427+0338,][]{OG2021}, for which archival and new optical spectroscopy, in conjunction with new X-ray observations, confirm the presence of a WD companion accreting material from the giant component.  
This makes it a bona fide X-SySt.  
Very remarkably, CGCS\,6306 is a carbon Mira, included in the {\it General Catalog of Galactic Carbon Stars} \citep{Alksnis2001} after the detection of C-rich features in its optical and infrared spectrum \citep{Cohen1996}. 
Therefore CGCS\,6306 provides an addition to the exclusive group of Galactic carbon SySts.

The paper is organized as follows.  
The archival and new optical spectroscopic and photometric, and X-ray observations of CGCS\,6306 are described in Sect.~\ref{sec.obs}.  
The results based on the analyses of these observations are presented in Sect.~\ref{sec.res}. 
The discussion and final remarks are provided in Sects.~\ref{sec.dis} and \ref{sec.fin}.

\section{Observations}
\label{sec.obs}

\subsection{X-ray observations}

CGCS\,6306 was serendipitously observed by XMM-Newton on 2010 April 30 using the European Photon Imaging Camera (EPIC) when targeting the cluster of galaxies MACS\,J0845+03 \citep[PI: S.~Allen, Obs.~ID.~0650381601;][]{Webb+2020}. 
The total exposure times of the EPIC pn, MOS1, and MOS2 cameras were 6.8, 8.4, and 8.4 ks, respectively. 
The data were processed using the Science Analysis Software \citep[SAS version 17.0;][]{Gabriel2004}. 
The event files were processed using the {\it epproc} and {\it emproc} SAS tasks. After excising bad periods of time, the net exposure times for the MOS1 and MOS2 cameras resulted in 4.9 and 4.6 ks, respectively. 
As for the EPIC pn camera, the bad period screening resulted in a not useful too short exposure time.  
A preliminary analysis of the XMM-Newton data on CGCS\,6306 was presented by \citet[][]{OG2021}.

The cluster of galaxies MACS\,J0845+03 has been recently targeted by the Chandra X-ray Observatory (CXO) using the Advanced CCD Imaging Spectrometer (ACIS) array on 2023 December 28 and 2024 January 5 (PI: A.D.~Goulding) with total exposure times of 11.95 and 15.54 ks corresponding to Obs.\ ID.\ 29163 and 26729, respectively.  
CGCS\,6306 is once again serendipitously registered in these observations located close to the aim point of the back-illuminated CCD S3. 
The CXO data were reprocessed with the Chandra Interactive Analysis of Observations \citep[CIAO, version 4.14,][]{Fruscione+2006} using the {\it chandra\_repro} task. 
No time periods of enhanced background emission were detected by inspecting background light curves in the 5.0–10.0 keV, which results in net exposure times equal to the total exposure times.

\begin{figure*}
\begin{center}
\includegraphics[width=1.66\columnwidth]{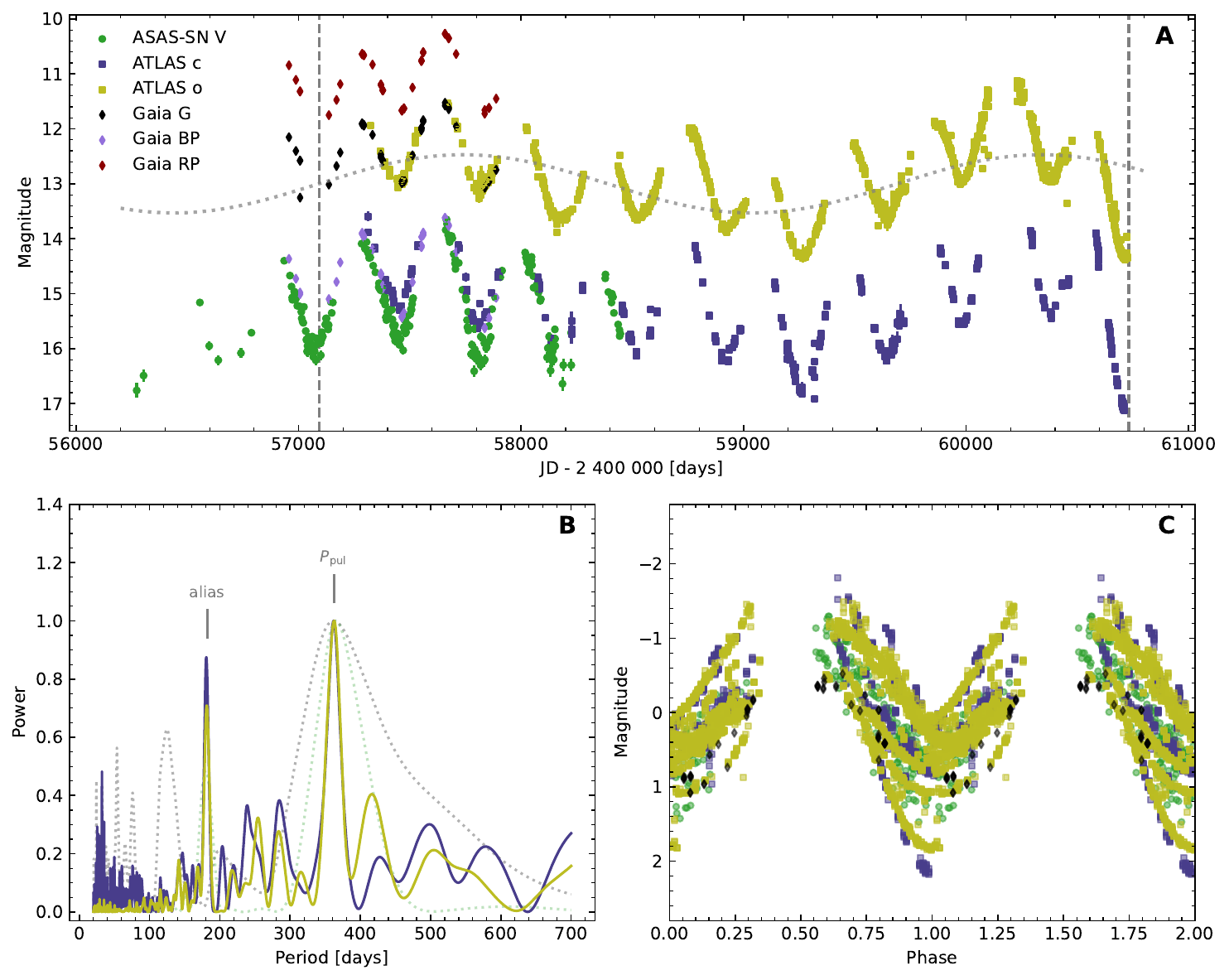}
\caption{
Photometry of CGCS\,6306. 
\textbf{A:} Light curves from ASAS-SN, ATLAS, and Gaia. 
The dotted gray sinusoidal line represents the apparent long-term variability trend. The vertical lines in the light curves mark the observation time of the LAMOST (2015 March 11) and NOT/FIES (2025 February 25) spectra.
\textbf{B:} Lomb-Scargle periodograms of the photometric data. 
Due to the short timespan of ASAS-SN and Gaia $G$ observations, their periodograms are shown as dotted lines only. 
\textbf{C:} Phase-folded light curves of CGCS\,6306. 
The long-term sinusoidal trend has been removed, and the median magnitude was subtracted for clarity.  
}
\label{lcs_all}
\end{center}
\end{figure*}

The new CXO observations, with higher spatial resolution than those obtained by XMM-Newton, confirm the presence of a point-source of X-ray emission at the location of CGCS\,6306 (left panel of Fig.~\ref{img.x}).  
The count rates at different energy bands are listed in Tab.~\ref{tbl.x} for the two CXO observations, including as well those derived from the XMM-Newton EPIC MOS1 and  MOS2 data sets for comparison. 
Its X-ray emission is revealed to be notably hard, with no counts below 2.0 keV, as also shown in the CXO color-composite picture in the right panel of Fig.~\ref{img.x}.

\subsection{Optical spectroscopy}

CGCS\,6306 was observed by the Large Sky Area Multi-Object Fiber Spectroscopic Telescope (LAMOST) survey \citep{2012RAA....12.1197C,2012RAA....12..723Z}, where it has designation LAMOST J084522.25+032711.5. 
The spectrum, with a resolution of $R~\sim~1800$, was obtained on 2015 March 11, near a minimum of the stellar light curve (Fig. \ref{lcs_all}), with a total exposure time of 4500\,s and is available in LAMOST DR4.

High-dispersion spectra of CGCS\,6306 were acquired using the FIbre-fed Echelle Spectrograph (FIES) at the Nordic Optical Telescope (NOT) of the Observatorio de El Roque de los Muchachos (ORM, La Palma, Spain) on 2025 February 25 at UT 23:42 to 00:36.  
The e2v Technologies back-illuminated CCD231-42 2048$\times$2064 CCD \#15 was used.  
Its pixel size of 15 $\mu$m, together with the low-resolution F1 fiber bundle used during these observations, provide a spectral dispersion of 0.0261 \AA~pixel$^{-1}$ and a spectral coverage from 3630 to 9270 \AA.  
The spectral resolution is $R \sim 25000$. 
Three 1040 s exposures were obtained under optimal sky transparency and seeing ($\simeq$1.0 arcsec) conditions. Like the LAMOST observations, the NOT FIES spectrum was also obtained at an epoch near a minimum of the stellar light curve (Fig. \ref{lcs_all}).

\subsection{Photometric observations}

Photometric data for CGCS\,6306 have been compiled from several surveys, including information in the $V$ and $g$ filters from the All-Sky Automated Survey for Supernovae \citep[ASAS-SN;][]{2014ApJ...788...48S, 2017PASP..129j4502K}, in the non-standard $c$ and $o$ filters from the Asteroid Terrestrial-impact Last Alert System\footnote{Data retrieved via the ATLAS Forced Photometry server \citep{2021TNSAN...7....1S}.} \citep[ATLAS; ][]{2018PASP..130f4505T,2020PASP..132h5002S}, and in the $G$, $BP$, and $RP$ bands of \textit{Gaia} \citep[\textit{Gaia} DR3; ][]{2023A&A...674A...1G}. 
The ASAS-SN observations in the $g$ band yielded mostly non-detections, so we excluded them from further analysis.

\begin{figure*}
\begin{center}
\includegraphics[width=1.7\columnwidth]{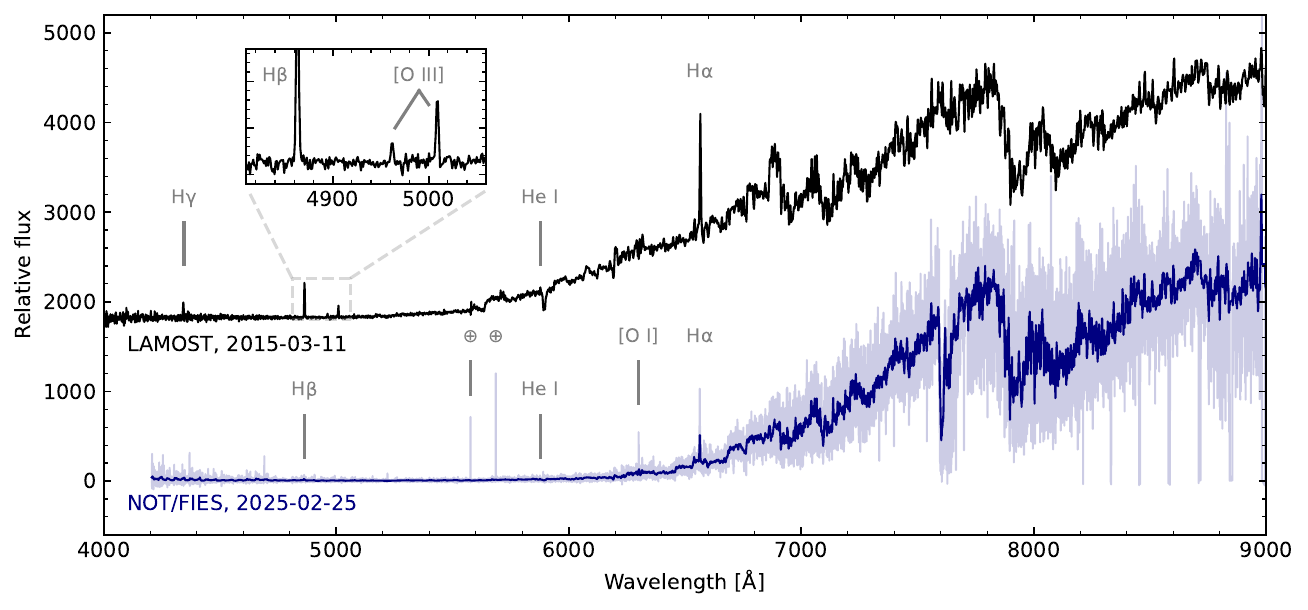}
\caption{LAMOST and NOT FIES optical spectra of CGCS\,6306. 
The spectra are shifted vertically by a constant.  
A running window median filter was applied to the NOT FIES spectrum for clarity. 
Detected prominent emission lines are marked by vertical lines. 
Telluric lines are marked as well.  
}

\label{spec.opt}
\end{center}
\end{figure*}

\begin{figure}
\begin{center}
\includegraphics[width=0.75\columnwidth]{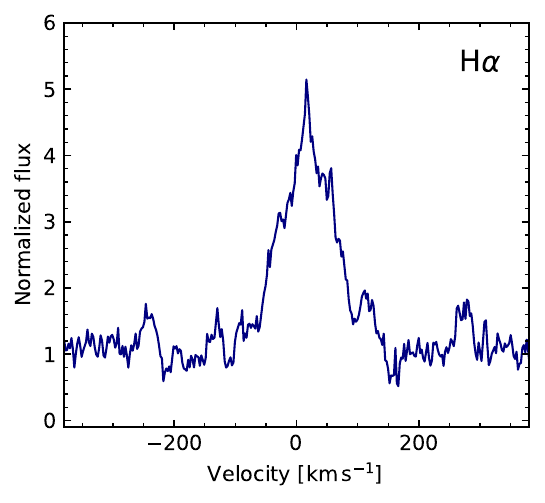}
\includegraphics[width=0.75\columnwidth]{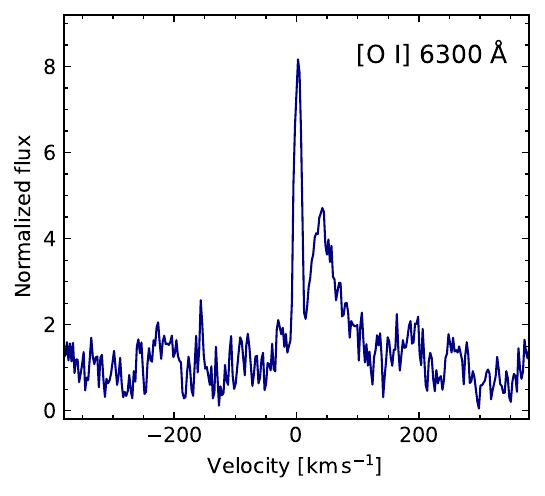}
\caption{
Zoom-in view of the H$\alpha$ (top) and [O~{\sc i}] $\lambda$6300 \AA\ (bottom) emission lines in the NOT FIES spectrum. 
Velocities are referred to the Heliocentric frame. 
The narrow blue component of the [O~{\sc i}] $\lambda$6300 \AA\ emission line at $\approx$0 km~s$^{-1}$ corresponds to a telluric contribution. 
}
\label{spec.opt.zoom}
\end{center}
\end{figure}

\section{Results}
\label{sec.res}

\subsection{Optical variability, luminosity, and distance}

CGCS\,6306 has been identified as a long-period variable (LPV) 0845224+032712 in the {\it Northern Sky Variability Survey} \citep[NSVS,][]{Wozniak2004a} with $P = 288~{\rm days}$ \citep{Wozniak2004b}.
Subsequent $I$ photometric monitoring indicated a longer period of $325~{\rm days}$ \citep{Mauron2014}, 
whereas even longer periods of $348.5~{\rm days}$ and $386~{\rm days}$ are listed in the ATLAS survey \citep{2018AJ....156..241H} and ASAS-SN catalog of variable stars \citep{2018MNRAS.477.3145J}, respectively. 
Finally, \citet{Lebzelter2023} used Gaia $G$ band data to obtain a period of $370\pm28~{\rm days}$.

We have used here ASAS-SN $V$, ATLAS $c$ and $o$, and Gaia $G$, $BP$ and $RP$ observations spanning about 12 yr (top panel of Fig.~\ref{lcs_all}) to investigate the optical variability of CGCS\,6303.  
To analyze the periodicity, we applied the Lomb-Scargle method \citep{1976Ap&SS..39..447L,1982ApJ...263..835S} using the {\tt astropy} Python package \citep{2013A&A...558A..33A,2018AJ....156..123A,2022ApJ...935..167A}.
The combined light curves display a short-term variability of $\simeq$2 mag overimposed on a long-term $\simeq$1 mag modulation.  
The latter implies a period $\simeq$7 yr, whereas our period analysis of ATLAS $c$ and $o$ observations led to a period of 361.9 $\pm$ 19.3 days and 362.6 $\pm$ 19.3 days, respectively (see panel B of Fig. \ref{lcs_all}). Analysis of ASAS-SN $V$ and \textit{Gaia} $G$ band data yielded similar results (365.3 $\pm$ 30.7 days and 362.0 $\pm$ 70.4 days, respectively). However, the uncertainties are significantly larger due to the much shorter timespan of the data. 
We adopted the average values from ATLAS $c$ and $o$,  
$362\pm14~{\rm days}$ as the final period.  
The large amplitude, almost sinusoidal variation, and long period of the light curve of CGCS\,6306 are typical of a pulsating Mira-type variable.

The origin of the longer, $\approx$7 yr periodicity is unclear. 
Many LPVs, mainly OGLE small-amplitude red giants (OSARGs) but also semi-regular variables (SRVs), exhibit so-called long secondary periods (LSPs). 
Although LSPs have been known for a long time, their origin remains uncertain. 
One proposed explanation is non-radial pulsations, but it is not yet clear whether the same phenomenon observed in OSARGs and SRVs is also present in Miras in general \citep[e.g.,][]{Pawlak+2023} and in CGCS\,6306 in particular.  
Alternatively LSPs have been attributed in some cases to binarity \citep[see, e.g.,][and references therein]{Wood+2004,Soszynski+2021,Pawlak+2024}, as it was also the case for Y\,Gem \citep{Guerrero+2025}. 
In SySts, orbital periods are often detectable in photometry, either due to the reflection effect and asymmetric nebulae, or ellipsoidal variability when the giant is (close to) filling its Roche lobe \citep[e.g.,][]{Munari+2019}. 
As for CGCS\,6306, the contribution of nebular emission does not appear to be strong enough to produce variability. 

If the observed variability were caused by tidal deformation of the giant, this would suggest an orbital period approximately twice as long ($\approx$14 years), but it would imply an unlikely large stellar radius, $\sim$1000~R$_\odot$, for a carbon giant \citep[with a typical radius 2--3$\times$ smaller; e.g.,][]{1996AJ....112..294D,1997AJ....114.2150V,2013ApJ...775...45V}, inconsistent with the inferred luminosity (see below). In addition, 
the observed magnitude amplitude, $\Delta m \approx 1$ mag, appears too large for ellipsoidal variability \citep[see, e.g., Fig. 10 in][]{2023A&A...680A..36G}.

Being CGCS\,6306 a pulsating Mira-type variable, its absolute bolometric magnitude and luminosity can be derived from period-luminosity ({\it P-L}) relationships derived for Galactic carbon Miras, such as those provided by \citet{GW1996} 
\begin{equation}
    M_{bol} = -2.59\log P + 2.02 = -4.61\,\mathrm{mag}\;(\sigma = 0.26), 
\end{equation}
or \citet{,Whitelock2006}
\begin{equation}
M_{bol} = -2.54 \log P + 1.87 = -4.63\,\mathrm{mag}\;(\sigma = 0.17).
\end{equation}
The average value implies a luminosity of $5.3 \times 10^3 L_{\odot}$. This result is similar to \citet{Lagadec2012}, who modeled the spectral energy distribution using fluxes derived from the near-IR photometry obtained by the {\it 2MASS} and {\it WISE} all-sky surveys \citep{Cutri2003,Wright2010} and the mid-infrared spectroscopy by {\it Spitzer}. 
The IR flux densities were fitted using the {\sc dusty} code \citep{Ivezic1999} to obtain a luminosity of $L = 7.5 \times 10^3 L_{\odot}$.  
A mass-loss rate and an expansion velocity of the circumstellar dust shell of $\dot{M}=4.2 \times 10^{-6}M_{\odot}$~yr$^{-1}$ and $V_{\rm exp}=16.5$ km~s$^{-1}$, respectively, are derived from CO ($J=3-2$) observations.

The distance towards CGCS\,6306 cannot be derived from similar {\it P-$M_K$} relationships because the 2MASS $K$ magnitude of $6.255\pm0.020$ mag was obtained near minimum, and the variation of the amplitude in this band is not known, but it is certainly large given the observed amplitude of variability of $\approx 1$ mag in the WISE W1 and W2 bands or reported for the $R$ and $I$ bands \citep{Wozniak2004b,Mauron2014}.  
Instead the distance can be derived from its \textit{Gaia} DR3 parallax, resulting in a value of $3.8\pm0.5$ kpc, which has been revised down to $3.4\pm0.4$ kpc by \citet{2021AJ....161..147B}. 
The height over the Galactic Plane of CGCS\,6306 ($l=223.49^{\circ}$, $b=+26.82^{\circ}$) would be $1.53\pm0.18$ kpc.  
The interstellar reddening along this direction, close to the galactic anti-center and high over the Galactic Plane, is very low: $E(B-V)=0.042$ \citep[or $A_V=0.13$ mag, assuming $R_V=3.1$,][]{Schlegel1998}. 

The position of CGCS\,6306 in the Galaxy, at the distance of 3.4 kpc and displaced 1.5 kpc from the Galactic plane, indicates that the star is possibly a member of the thick disc \citep{Lagadec2012} or the halo population \citep{Mauron2007,Mauron2008}. 
According to \citet{Mauron2014} the population of C-rich AGB stars in the halo could be the result of the disruption of satellite dwarf galaxies, such as the dwarf Sgr, for example.

\subsection{Optical spectra}

The low-resolution spectrum of CGCS\,6306 presented by \citet{Cohen1996} enabled its classification as a carbon star of spectral type C6,4. 
The authors also reported H$\alpha$ emission\footnote{
Their figure 4 reveals the presence of an H$\beta$ emission line as well. 
}
and a flat blue spectral shape, which they suggested was reminiscent of a hot companion.

The archival LAMOST and our NOT FIES spectra of CGCS\,6306 (Fig.~\ref{spec.opt}) display a number of emission lines in addition to the H$\beta$ and H$\alpha$ H{\sc i} Balmer lines overimposed on a late-type stellar continuum.  
The emission lines identified in the LAMOST spectrum are [O~{\sc iii}] $\lambda\lambda$4959,5007 \AA\ and He~{\sc i} $\lambda$5876 \AA, whereas those in the NOT FIES spectrum are He~{\sc i} $\lambda$5876 \AA\ and [O~{\sc i}] $\lambda\lambda$6300,6363 \AA.

Compared to the stellar continuum, the intensity of the Balmer lines is notably brighter on the LAMOST spectrum than on the NOT FIES spectrum 10 years later.  
Otherwise, the different emission lines detected in the LAMOST and NOT FIES spectra do not necessarily mean variability;  
the [O~{\sc iii}] emission lines detected by LAMOST are too faint to be detected in the NOT FIES spectrum, whereas the detection [O~{\sc i}] emission lines in the latter spectrum has been allowed by its high dispersion, which deblends the stellar and telluric contributions of these lines.  
We emphasize that the detection of these faint nebular lines has been favored by the acquisition of the LAMOST and NOT FIES optical spectra at epochs near the minimum of the stellar light curve.

The high-dispersion NOT FIES spectrum reveals that the H$\alpha$ emission line profile (Fig.~\ref{spec.opt.zoom}-top) does not show a central absorption \citep{IT2004} nor a P Cygni profile \citep{Izumiura2008,McKeever2011}, unlike most SySts, but it resembles a Gaussian curve showing wings up to $\approx 120$ km~s$^{-1}$.  
The [O~{\sc i}] emission lines are wide as well (Fig.~\ref{spec.opt.zoom}-bottom), with a FWHM of 70 km~s$^{-1}$ compared to that of 110 km~s$^{-1}$ for H$\alpha$.  
The H$\alpha$ profile can be interpreted as the contribution of two components at velocities $\pm$65 km~s$^{-1}$ according to expression 
\begin{equation}
    V = \frac{FWHM}{2 \sqrt{\ln 2}} = 0.60 \times FWHM. 
\end{equation}

\subsection{X-ray spectroscopy}
\label{sec:xray}

\begin{table*}
\begin{center}
\caption{Results of the spectral analysis performed with XSPEC.}
\tiny
\setlength{\tabcolsep}{\tabcolsep}  
\begin{tabular}{llcccccc} 
\hline
Epoch & Instrument & $\chi^{2}_\mathrm{DoF}$ & $N_\mathrm{H}$ & $kT$  & $A$       & $f_\mathrm{X}$  & 
$F_\mathrm{X}$ \\
       &     &      & (10$^{21}$ cm$^{-2}$)    & (keV) & (10$^{-5}$ cm$^{-5}$) & (erg~cm$^{-2}$~s$^{-1}$)            & (erg~cm$^{-2}$~s$^{-1}$)\\
\hline
2010-04-30 & XMM MOS(1+2)& 1.06 & {\bf 9.0}     & {\bf 10.0}   & $(2.9\pm1.0)\times10^{-4}$ & $(2.5\pm0.9)\times10^{-13}$ & $(6.2\pm2.1)\times10^{-13}$\\
2023-12-28 & CXO ACIS-S & 1.44  & 30$\pm$16 &  {\bf 10.0}  & $(1.0\pm0.8)\times10^{-4}$ & $(4.8\pm3.5)\times10^{-14}$ & $(2.2\pm1.6)\times10^{-13}$\\
2024-01-05 & CXO ACIS-S & 1.85  & 8.9$\pm$3.4   &  {\bf 10.0}  & $(7.8\pm2.6)\times10^{-5}$ & $(6.8\pm2.2)\times10^{-14}$ & $(1.7\pm0.5)\times10^{-13}$ \\
\hline
\end{tabular}
\label{tab:xray}
\tablefoot{
Observed ($f_\mathrm{X}$) and intrinsic ($F_\mathrm{X}$) fluxes computed in the 0.3--10.0 keV energy range. 
The normalization parameter ($A$) is defined as $A = 10^{-14}\int n_\mathrm{H} n_\mathrm{e} dV / 4 \pi d^{2}$, where $n_\mathrm{H}$ and $n_\mathrm{e}$ are the hydrogen and electron densities, $V$ is the volume of the X-ray-emitting region, and $d$ is the distance. Boldface values represent fixed values during the spectral analysis.
}
\end{center}
\end{table*}

\begin{figure}
\begin{center}
\includegraphics[width=0.94\columnwidth]{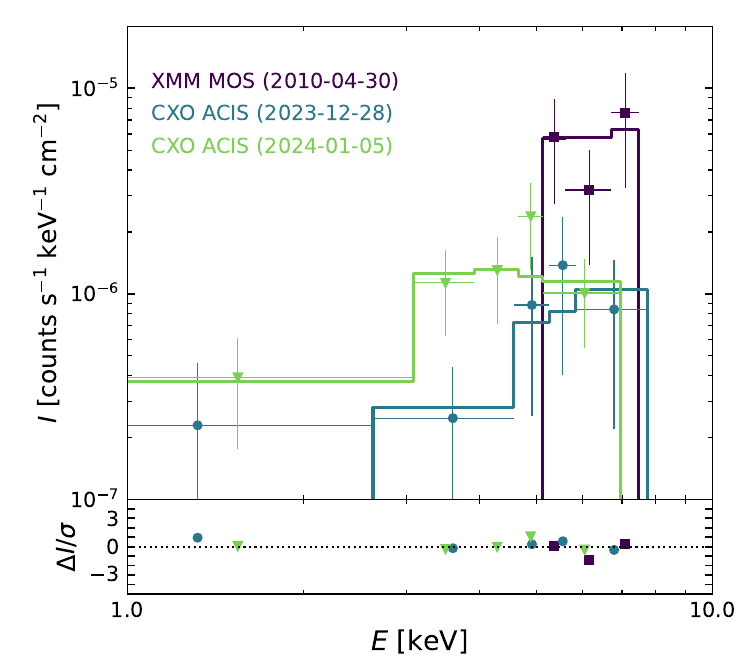}
\end{center}
\caption{Background-subtracted X-ray spectra of CGCS 6306. Different symbols show the spectra from different epochs. The solid lines show the best-fit models described in Table~\ref{tab:xray}. Note that the spectrum labeled as MOS is the combination of the MOS1 and MOS2 spectra.}
\label{spec.x}
\end{figure}

The spectra extraction region from the CXO ACIS S3 observations was performed from a circular aperture with a radius of 3.0\arcsec{} centered on CGCS~6306. The background spectrum was selected from a set of larger circular regions free from background sources around CGCS\,6306 and with similar distances to the core of MACS\,J0845+03 to obtain a fair representation of its diffuse emission. The source and background spectra were produced by making use of the {\it specextract} CIAO task, which simultaneously produces all necessary calibration files. The background-subtracted CXO ACIS spectra of CGCS\,6306 derived from the 2023.99 and 2024.01 observations are presented in Fig.~\ref{spec.x}.

Given the larger PSF of XMM-Newton EPIC instruments (PSF\,$\sim 6''$), compared to that of the CXO ACIS  (PSF\,$\sim1''$), we extracted the spectrum of CGCS~6306 using a circular aperture of 13\arcsec{} in radius. The background was selected following a similar approach as that for the CXO observations. Spectra from the two MOS cameras were extracted using the SAS task {\it evselect} and their associated calibration files were produced using the {\it rmfgen} and {\it arfgen} tasks. Finally, a single MOS spectrum was produced by combining the MOS1 and MOS2 spectra with the SAS task {\it epicspeccombine}. The combined background-subtracted MOS spectrum is plotted in Fig.~\ref{spec.x} alongside those obtained from the CXO observations.

Fig.~\ref{spec.x} shows that the X-ray spectra of CGCS~6306 have been consistently hard for the past 14 yr, with the bulk emission peaking at $E>$ 5 keV and an absent contribution at softer energies. 
The latter is suggestive of a highly-extinguished hard X-ray emission. To assess the physical properties of the X-ray emission detected from CGCS~6306, we have modeled the X-ray spectra using the X-Ray Spectral Fitting Package \citep[XSPEC; version 12.12.1;][]{Arnaud1996}. An optically-thin plasma emitting model {\it apec}\footnote{\url{https://heasarc.gsfc.nasa.gov/xanadu/xspec/manual/node134.html}} was adopted with Solar abundances \citep{Anders1989}. The extinction was considered making use of the {\it tbabs} component included in XSPEC \citep[][]{Wilms2000}. Consequently, each spectrum is characterized by a plasma temperature $kT$ and an averaged hydrogen column density $N_\mathrm{H}$. The quality of the model is evaluated through the reduced $\chi^2$ statistics ($\chi^2_\mathrm{DoF}$) directly provided by XSPEC.

The results of the spectral analysis are summarized in Table~\ref{tab:xray}. 
The extinction was better characterized in the CXO spectrum from 2024.01, given its slightly superior count rate compared to the other two spectra. 
Given the limited count number, the plasma temperature was fixed at 10 keV. 
According to the spectral modeling, the flux of the two CXO epochs is consistent. 
The estimated intrinsic flux from the XMM MOS data seems a bit larger, but it is still marginally consistent with that estimated for the CXO data from 2023.99 within error values. Given the quality of the XMM data, we can safely argue that the X-ray flux seems to have remained constant. 
Adopting a constant flux of $F_\mathrm{X} \approx (2\pm1)\times10^{-13}$ erg cm$^{-2}$ s$^{-1}$, the X-ray luminosity can thus be estimated to be $L_\mathrm{X} \approx (2.8\pm1.4)\times10^{32}$ erg s$^{-1}$, for an assumed distance of $d=3.4$ kpc. 

\section{Discussion}
\label{sec.dis}

\subsection{Optical spectrum}\label{sec:opt_spec}

The optical spectra of AGB stars may present Fe~{\sc ii} emission lines irrespective of its Mira, SR- and L-type class of variability, but the H~{\sc i} Balmer series appear in emission only in the spectrum of Mira-type variables.  
The origin of both Fe~{\sc ii} and H~{\sc i} lines can be attributed to shocks propagating in the star's upper atmosphere \citep{Fox1984,RW2001}, but their intensities are highly variable along the Mira pulsation cycle: 
they are practically absent at stellar minimum and increase in intensity when the star approaches the maximum of its light curve, remaining strong until it completes one-third of its period \citep{GO2020}. 
Thus, the acquisition of spectra of suspected SySts at the minimum of the pulsation cycle of the giant component does not only favor the detection of faint nebular lines, such as those of [O~{\sc i}] and [O~{\sc iii}] in CGCS\,6306, which might become completely outshine at maximum light, but it also discards the origin of Balmer lines in the giant photosphere. 
This supports the classification of CGCS\,6306 as a bona fide SySt, as its optical spectra fulfill the traditional criteria for SySt classification, as outlined in the Introduction. Moreover, the position of the star in the [O~{\sc iii}] diagnostic diagram \citep{1995PASP..107..462G,2017A&A...606A.110I} further confirms its symbiotic nature ([O~{\sc iii}] $\lambda$4363/H$\gamma$ = 0.30 and [O~{\sc iii}] $\lambda$5007/H$\beta$ = 0.32).

The emission lines observed in the spectrum of CGCS\,6306 (H$\alpha$, H$\beta$, He{\sc i} $\lambda$5876 and [O~{\sc i}] $\lambda\lambda$6300,6363) show some similarities with those observed in the spectrum of RS\,Oph \citep[i.e. the Balmer series, He~{\sc i} $\lambda$5876 and {[O~{\sc i}]} $\lambda$6300,][]{Brandi2009}, a recurrent symbiotic nova star, during its quiescence phase. 
Recombination lines such as the Balmer series are generally formed in SySts close to the secondary star, which acts as a photoionization source \citep{Contini1997}. 
The high electron density in this region ($N_\mathrm{e} > 10^6$ cm$^{-3}$) would suppress the formation of some forbidden lines, which are generally formed farther away from the secondary, where the density is lower \citep[between $10^4$ and $10^6$ cm$^{-3}$,][]{ATP2003}. 
However, CGCS\,6306 unexpectedly shows broad [O~{\sc i}] $\lambda\lambda$6300,6363 profiles, with $W_{\rm FWHM} \approx 70$ km s$^{-1}$, which is comparable to the width of the H$\alpha$ line, $W_{\rm FWHM} \approx 110$ km s$^{-1}$ (Fig.~\ref{spec.opt.zoom}). 
Therefore, differently from most SySts, the [O~{\sc i}] doublet seems to be formed in a relatively dense gas, with $N_\mathrm{e} \simeq 10^6$ cm$^{-3}$. 
Actually, the critical density of this doublet is one of the highest among the nebular lines commonly observed in SySts and PNe \citep[$N_{\rm crit}$ = $2 \times 10^6$ cm$^{-3}$,][]{AO1988}. On the other hand, the spectral resolution of the available LAMOST spectrum detecting the [O~{\sc iii}] emission lines is not sufficient to determine their width, and thus to assess their possible origin in the accretion disk.  
Nevertheless, the extremely low ratio of the [O~{\sc iii}] $\lambda$5007 nebular line to the [O~{\sc iii}] $\lambda$4363 auroral line, $\approx$2.4, is indicative of a high density.
Although this cannot be firmly assessed due to the unknown extinction and electron temperature, reasonable assumptions suggest that the density is at least five times higher than the critical density of $6.3 \times 10^5$ cm$^{-3}$ for the [O~{\sc iii}] $\lambda\lambda$4959,5007 emission lines.  
The high density of the emitting region is further supported by the non-detection of the [S~{\sc ii}] $\lambda\lambda$6716,6731 doublet, which have critical densities of $1.6 \times 10^3$ and $4.0 \times 10^3$ cm$^{-3}$, respectively.

If we assume an accretion disk with Keplerian rotation, the velocity of material rotating in a circular orbit at radius $R$ around the secondary star with mass $M$ is described \citep{ATP2003,Brandi2009} as 

\begin{equation}
V = 437 \sqrt{\frac{M}{R}},
\end{equation}

\noindent where $V$ is in km s$^{-1}$, and $M$ and $R$ are given in solar units. 
Assuming a typical mass of a white dwarf $M = 0.6~M_{\odot}$ and adopting a rotation velocity of 90 km s$^{-1}$ correcting for an inclination angle of 45$^\circ$ the velocity derived from the FWHM of the H$\alpha$ line profile, a disk radius of $14\,R_{\odot}$ (= 0.065 AU = $1.0 \times 10^{12}$ cm) is obtained.
This is most likely a radius representative of the bulk of the H$\alpha$ emitting material in the disk.

\subsection{X-ray properties}

The SySt nature of CGCS\,6306 and the X-ray properties of its spectrum classify it as a $\delta$-type X-SySt under the scheme originally proposed by \citet{Murset1997} later extended by \citet{2013A&A...559A...6L}. 
These systems are characterized by a heavily-extinguished plasma with a hard X-ray-emitting spectrum.

The X-ray luminosity of CGCS\,6306 is also typical of $\delta$- and $\beta/\delta$-type X-ray-emitting SySts, of the order of $10^{32}$~erg~s$^{-1}$ \citep[see figure 5 of][]{Guerrero+2024}. 
If the X-ray emission is the result of accretion, \citet{Guerrero+2024} estimated mass accretion rate for those X-ray luminosities of $\dot{M}_\mathrm{acc} \gtrsim 10^{-11}$ M$_\odot$ yr$^{-1}$, adopting a typical radius and mass for the accreting WD of 0.01 R$_\odot$ and 0.6 M$_\odot$, respectively.
However, this estimate is solely based on the X-ray luminosity, and it does not account for the disk luminosity that is preferentially detected in UV and in the bluest optical regions. \citet{Cohen1996} noted a possible contribution to the spectrum of CGCS\,6306 in the blue region of its optical spectrum, whereas the GALEX NUV absolute magnitude ($\simeq$10 mag at the distance of 1.53 kpc) seems to imply UV emission excess \citep{OG2016}, although it cannot be confirmed since there are no available GALEX FUV observations of CGCS\,6306.
Thus the present estimate of the accretion rate onto the WD companion of CGCS\,6306 should be regarded as a lower limit.

\subsection{CGCS\,6306 as a rare carbon Mira SySt}

CGCS\,6306 belongs to a small group of carbon-rich SySts in the Milky Way.
Among the known Galactic carbon SySts, five are reported to host Mira-type pulsating components. 
The giant in V1196 Sco \citep[=AS 210, with a pulsation period $P_\mathrm{Mira} = 423$ day,][]{1950ApJ...112...72M,1967ApJS...14..125H,1978MNRAS.184..601A}, 
SS73\,38 \citep[$P_\mathrm{Mira} = 463$ day,][]{1973ApJ...185..899S,1987PASP...99..573W,2009AcA....59..169G}, 
H1-45 \citep[$P_\mathrm{Mira} = 409$ day,][]{Miszalski+2013}, UV\,Aur\footnote{The symbiotic nature of UV Aur remains debated, as \citet{Herbig2009} did not detect previously reported high-excitation emission lines. However, since these observations were obtained at pulsational maximum, it is possible that the emission lines were temporarily obscured, as further discussed in Section \ref{sec:opt_spec}.} \citep[$P_\mathrm{Mira} = 394$ day,][]{1950ApJ...111..270S,1988ASSL..145..293S,2013A&A...559A...6L,1995Ap&SS.225..101G}, and IPHAS J205836.43+503307.2 \citep{2011A&A...529A..56C}, which has been suggested as a likely member of this group but its pulsation period has not yet been reported.
In addition to these confirmed cases, several SySt candidates are listed in the New Online Database of Symbiotic Variables, including 
IRAS\,18120+4530, a carbon Mira \citep[$P_\mathrm{Mira} = 392$ day,][]{2014Ap.....57..510G} that exhibits H$\alpha$, H$\beta$, and [O~{\sc iii}] emission lines in its optical spectrum \citep{Mauron2008}, and 
OGLE-BLG-LPV-145820, which shows H$\alpha$, [O~{\sc ii}], and He~{\sc i} in its spectrum, although the available spectral coverage is limited to the 5600–9200\AA{} range \citep[$P_\mathrm{Mira} = 600$ day,][]{2017MNRAS.469.4949M}. 
Another system frequently mentioned in the literature as a possible SySt is V335\,Vul, a carbon Mira \citep[$P_\mathrm{Mira} = 347$ day,][]{2003ASPC..303..171S} whose symbiotic nature was proposed by \citet{1999IBVS.4668....1M} after detecting Balmer lines in its optical spectrum and an apparent outburst. 
However, \citet{2003ASPC..303..171S} suggested that the observed variability was due to a pulsational maximum rather than an actual outburst, leaving the symbiotic classification of V335 Vul uncertain.

As currently listed in the New Online Database of Symbiotic Variables, there are only ten confirmed carbon-rich SySts in the Milky Way. 
This constitutes a small fraction of the $\sim$300 known Galactic symbiotic stars \citep[][]{2019RNAAS...3...28M,2019AN....340..598M,2019ApJS..240...21A}, in contrast to the Large and Small Magellanic Clouds, where carbon SySts are more prevalent. 
This difference is probably linked to the lower overall metallicity of the Magellanic Clouds, which facilitates more efficient third dredge-up episodes during the asymptotic giant branch phase, allowing stars to develop carbon-rich atmospheres more easily \citep{Karakas2002,ML2003,2021A&A...650A.118S}.

\section{Final remarks}
\label{sec.fin}

The X-ray emission from CGCS\,6306 \citep{GO2020}, as that of many other X-AGB stars \citep{Guerrero+2024}, makes it a SySt candidate.  
As for the case of Y\,Gem \citep{Guerrero+2025}, the acquisition of optical spectroscopic observations of CGCS\,6306 has revealed the presence of high-excitation emission lines that disclose a nebula ionized by a hot component.  
Archival long-term photometric observations confirm a Mira-type pulsation period of 362 days, making CGCS\,6306 one of the very few Galactic carbon Mira SySts. 
There is another long-term period $\approx$ 7 yr present in the light curve, but its possible association with an orbital origin is unknown.  
Follow-up observations are necessary to confirm this link.

The X-ray properties of CGCS\,6306, with a hard spectrum typical of highly-extincted hot plasma emission and an X-ray luminosity $\approx10^{32}$ erg~s$^{-1}$, are consistent with those of $\delta$-type X-SySts.  
Its X-ray luminosity ($\approx$0.07 L$_\odot$) is actually comparable to that of Y\,Gem ($\approx$0.12 L$_\odot$).  
The broad H$\alpha$ profile in CGCS\,6306 has been used to derive an accretion disk characteristic radius $\approx$14~R$_\odot$, while its X-ray luminosity has allowed us to determine a lower limit to the accretion rate of $\gtrsim10^{-11}$ M$_\odot$~yr$^{-1}$.  
This accretion rate is about 20,000 times smaller that inferred for Y\,Gem, but we note that the accretion rate in Y\,Gem is mostly based on the luminosity of its accretion disk, which is $\gtrsim$200 times larger than its X-ray luminosity.   
Unfortunately the contribution to the accretion rate from the luminosity of the accretion disk of CGCS\,6306 is unknown, since it ought to be derived from UV and blue optical observations, which are not available or sufficiently restrictive.

We would like to emphasize the advantages that the acquisition of optical spectroscopic observations carried out at an epoch near the light curve minimum of an X-AGB star have on the assessment of its possible SySt nature.  
At this phase of the pulsational cycle, some emission features originated in the atmosphere of the AGB star, such as H~{\sc i} recombination lines, are not expected to show, strengthening the case for its association with accretion processes (see the inconclusive case of the spectroscopic observations at or near maximum of UV\,Aur and V335\,Vul described in Sect.~4.3).
Furthermore, the higher contrast of nebular emission lines with respect to a diminished stellar continuum at minimum makes their detection more likely, which provides conclusive evidence of the SySt nature.

\begin{acknowledgements}

We thank the anonymous referee for their helpful comments that improved the manuscript. We thank Shreeya Shetye for the fruitful discussion on the third dredge-up in AGB stars.  
This work is based on service observations made with the Nordic Optical Telescope (programme SST2025-681; PI: M. A. Guerrero), owned in collaboration by the University of Turku and Aarhus University, and operated jointly by Aarhus University, the University of Turku and the University of Oslo, representing Denmark, Finland and Norway, the University of Iceland and Stockholm University at the Observatorio del Roque de los Muchachos, La Palma, Spain, of the Instituto de Astrofísica de Canarias. 
The contact person, Rosa Clavero, the astronomers who conducted the observations, David Jones and Paulina Sowicka, and the FIES instrument specialist, John Telting, are particularly appreciated for their guidance and support.  
This work is based on observations obtained with XMM-Newton, an European Science Agency (ESA) science mission with instruments and contributions directly funded by ESA Member States and NASA, and by the Chandra X-ray Observatory. 
The research of J.M.\ was supported by the Czech Science Foundation (GACR) project no. 24-10608O. 
M.A.G.\ acknowledges financial support from grants CEX2021-001131-S funded by MCIN/AEI/10.13039/501100011033 and PID2022-142925NB-I00 from the Spanish Ministerio de Ciencia, Innovaci\'on y Universidades (MCIU) cofunded with FEDER funds. 
J.A.T.\ acknowledges support from the UNAM PAPIIT project IN~102324.  
R.O.\ thanks the support of the São Paulo Research Foundation (FAPESP), grant \#2023/05298-0.

This work has made extensive use of NASA’s Astrophysics Data System.

\end{acknowledgements}

%%%%%%%%%%%%%%%%%%%% REFERENCES %%%%%%%%%%%%%%%%%%

% WARNING
%-------------------------------------------------------------------
% Please note that we have included the references to the file aa.dem in
% order to compile it, but we ask you to:
%
% - use BibTeX with the regular commands:
%   \bibliographystyle{aa} % style aa.bst
%   \bibliography{Yourfile} % your references Yourfile.bib

\begin{thebibliography}{}

\bibitem[Akras et al.(2019)]{2019ApJS..240...21A} Akras, S., Guzman-Ramirez, L., Leal-Ferreira, M.~L., et al.\ 2019, \apjs, 240, 21. %doi:10.3847/1538-4365/aaf88c

\bibitem[\protect\citeauthoryear{Alksnis et al.}{2001}]{Alksnis2001}
Alksnis, A., Balklavs, A., Dzervitis, U. et al., 2001, Baltic Astronomy, 10, 1

\bibitem[Allen(1978)]{1978MNRAS.184..601A} Allen, D.~A.\ 1978, \mnras, 184, 601. %doi:10.1093/mnras/184.3.601

\bibitem[Anders \& Grevesse(1989)]{Anders1989} Anders, E. \& Grevesse, N.\ 1989, \gca, 53, 197. %doi:10.1016/0016-7037(89)90286-X

\bibitem[\protect\citeauthoryear{Appenzeller \& \"Ostreicher}{1988}]{AO1988}Appenzeller, I. \& \"Ostreicher, R., 1988, AJ, 95, 45

\bibitem[Arnaud(1996)]{Arnaud1996} Arnaud, K.~A.\ 1996, Astronomical Data Analysis Software and Systems V, 101, 17

\bibitem[\protect\citeauthoryear{Arrieta \& Torres-Peimbert}{2003}]{ATP2003}Arrieta, A. \& Torres-Peimbert, S., 2003, ApJS, 147, 97

\bibitem[Astropy Collaboration et al.(2013)]{2013A&A...558A..33A} Astropy Collaboration, Robitaille, T.~P., Tollerud, E.~J., et al.\ 2013, \aap, 558, A33. %doi:10.1051/0004-6361/201322068
\bibitem[Astropy Collaboration et al.(2018)]{2018AJ....156..123A} Astropy Collaboration, Price-Whelan, A.~M., Sip{\H{o}}cz, B.~M., et al.\ 2018, \aj, 156, 123. %doi:10.3847/1538-3881/aabc4f
\bibitem[Astropy Collaboration et al.(2022)]{2022ApJ...935..167A} Astropy Collaboration, Price-Whelan, A.~M., Lim, P.~L., et al.\ 2022, \apj, 935, 167. %doi:10.3847/1538-4357/ac7c74

\bibitem[\protect\citeauthoryear{Ayres et al.}{1981}]{Ayres1981}
Ayres, T.R., Linsky, J.L., Vaiana, G.S., Golub, L. \& Rosner, R., 1981, \apj, 250, 293

\bibitem[Bailer-Jones et al.(2021)]{2021AJ....161..147B} Bailer-Jones, C.~A.~L., Rybizki, J., Fouesneau, M., et al.\ 2021, \aj, 161, 147. %doi:10.3847/1538-3881/abd806

\bibitem[Belczy{\'n}ski et al.(2000)]{Belczynski+2000} 
Belczy{\'n}ski, K., Miko{\l}ajewska, J., Munari, U., et al.\ 2000, \aaps, 146, 407. 
% doi:10.1051/aas:2000280

\bibitem[\protect\citeauthoryear{Brandi et al.}{2009}]{Brandi2009}Brandi, E., Quiroga, C., Mikolajewska, J., Ferrer, O.E. \& Garc\'\i a, L.G., 2009, A\&A, 497, 815


\bibitem[\protect\citeauthoryear{Cohen et al.}{1996}]{Cohen1996} Cohen, M., Wainscoat, R.J.,, Walker, H.J. \& Volk, K., 1996, \aj, 111, 1333

\bibitem[\protect\citeauthoryear{Contini}{1997}]{Contini1997}Contini, M., 1997, ApJ, 483, 887

\bibitem[Corradi et al.(2011)]{2011A&A...529A..56C} Corradi, R.~L.~M., Sabin, L., Munari, U., et al.\ 2011, \aap, 529, A56. %doi:10.1051/0004-6361/201016406

\bibitem[Cui et al.(2012)]{2012RAA....12.1197C} Cui, X.-Q., Zhao, Y.-H., Chu, Y.-Q., et al.\ 2012, Research in Astronomy and Astrophysics, 12, 1197. doi:10.1088/1674-4527/12/9/003

\bibitem[\protect\citeauthoryear{Cutri et al.}{2003}]{Cutri2003}
Cutri et al., 2003, {\it The IRSA 2MASS All-Sky Point Source Catalog}, NASA/IPAC Infrared Science Archive.

\bibitem[Dyck et al.(1996)]{1996AJ....112..294D} Dyck, H.~M., van Belle, G.~T., \& Benson, J.~A.\ 1996, \aj, 112, 294. %doi:10.1086/118014


\bibitem[\protect\citeauthoryear{Fox et al.}{1984}]{Fox1984}Fox, M.W., Wood, P.R. \& Dopita, M.A., 1984, ApJ, 286, 337

\bibitem[Fruscione et al.(2006)]{Fruscione+2006} 
Fruscione, A., McDowell, J.~C., Allen, G.~E., et al.\ 2006, \procspie, 6270, 62701V. 
% doi:10.1117/12.671760

\bibitem[Gabriel et al.(2004)]{Gabriel2004} Gabriel, C., Denby, M., Fyfe, D.~J., et al.\ 2004, Astronomical Data Analysis Software and Systems (ADASS) XIII, 314, 759

\bibitem[Gaia Collaboration et al.(2023)]{2023A&A...674A...1G} Gaia Collaboration, Vallenari, A., Brown, A.~G.~A., et al.\ 2023, \aap, 674, A1.% doi:10.1051/0004-6361/202243940

\bibitem[Gaia Collaboration et al.(2023)]{2023A&A...680A..36G} Gaia Collaboration, Trabucchi, M., Mowlavi, N., et al.\ 2023, \aap, 680, A36. %doi:10.1051/0004-6361/202347287



\bibitem[Gal \& Szatmary(1995)]{1995Ap&SS.225..101G} Gal, J. \& Szatmary, K.\ 1995, \apss, 225, 101. %doi:10.1007/BF00657847

\bibitem[Gigoyan et al.(2014)]{2014Ap.....57..510G} Gigoyan, K.~S., Sarkissian, A., Russeil, D., et al.\ 2014, Astrophysics, 57, 510. %doi:10.1007/s10511-014-9354-5

\bibitem[\protect\citeauthoryear{Groenewegen \& Whitelock}{1996}]{GW1996}
Groenewegen, M.A.T. \& Whitelock, P.A., 1996, \mnras, 281, 1347

\bibitem[Gromadzki et al.(2009)]{2009AcA....59..169G} Gromadzki, M., Miko{\l}ajewska, J., Whitelock, P., et al.\ 2009, \actaa, 59, 169. %doi:10.48550/arXiv.0906.4136

\bibitem[\protect\citeauthoryear{Guerrero \& Ortiz}{2020}]{GO2020}
Guerrero, M.A. \& Ortiz, R., 2020, \mnras, 491, 680

\bibitem[\protect\citeauthoryear{Guerrero et al.}{2024}]{Guerrero+2024} 
Guerrero, M.A., Montez, R., Ortiz, R., Toal\'a, J.A. \& Kastner, J.H., 2024, \aap, 689, A62

\bibitem[Guerrero et al.(2025)]{Guerrero+2025} 
Guerrero, M.~A., Vasquez-Torres, D.~A., Rodr{\'\i}guez-Gonz{\'a}lez, J.~B., et al.\ 2025, \aap, 693, A203. 
% doi:10.1051/0004-6361/202451715

\bibitem[Gutierrez-Moreno et al.(1995)]{1995PASP..107..462G} Gutierrez-Moreno, A., Moreno, H., \& Cortes, G.\ 1995, \pasp, 107, 462. %doi:10.1086/133575


\bibitem[Heinze et al.(2018)]{2018AJ....156..241H} Heinze, A.~N., Tonry, J.~L., Denneau, L., et al.\ 2018, \aj, 156, 241. %doi:10.3847/1538-3881/aae47f

\bibitem[Henize(1967)]{1967ApJS...14..125H} Henize, K.~G.\ 1967, \apjs, 14, 125. %doi:10.1086/190151

\bibitem[\protect\citeauthoryear{Herbig}{2009}]{Herbig2009}
Herbig, G.H., 2009, \aj, 138, 1502

\bibitem[Hynes et al.(2014)]{2014ApJ...780...11H} 
Hynes, R.~I., Torres, M.~A.~P., Heinke, C.~O., et al.\ 2014, \apj, 780, 11. %doi:10.1088/0004-637X/780/1/11

\bibitem[\protect\citeauthoryear{Iketa \& Tamura}{2004}]{IT2004}Ikeda, Y. \& Tamura, S., 2004, PASJ, 56, 353

\bibitem[I{\l}kiewicz \& Miko{\l}ajewska(2017)]{2017A&A...606A.110I} I{\l}kiewicz, K. \& Miko{\l}ajewska, J.\ 2017, \aap, 606, A110. %doi:10.1051/0004-6361/201731497


\bibitem[\protect\citeauthoryear{Ivezic et al.}{1999}]{Ivezic1999}
Ivezic, Z., Nenkova, M. \& Elitzur, M., 1999, User Manual for {\sc dusty} , University Kentucky Internal Report

\bibitem[\protect\citeauthoryear{Izumiura et al.}{2008}]{Izumiura2008}Izumiura, H., Noguchi, K., Aoki, W., Honda, S., Ando, H., Takada-Hidai, M., Kambe, E., Kawanomoto, S.,
Sadakane, K., Sato, B. et al., ApJ, 682, 499

\bibitem[Jayasinghe et al.(2018)]{2018MNRAS.477.3145J} Jayasinghe, T., Kochanek, C.~S., Stanek, K.~Z., et al.\ 2018, \mnras, 477, 3145. %doi:10.1093/mnras/sty838

\bibitem[\protect\citeauthoryear{Karakas et al.}{2002}]{Karakas2002}Karakas, A.I., Latanzzio, J.C. \& Pols, O.R., 2002, PASA, 19, 515

\bibitem[Kochanek et al.(2017)]{2017PASP..129j4502K} Kochanek, C.~S., Shappee, B.~J., Stanek, K.~Z., et al.\ 2017, \pasp, 129, 104502. %doi:10.1088/1538-3873/aa80d9

\bibitem[\protect\citeauthoryear{Lagadec et al.}{2012}]{Lagadec2012}
Lagadec, E., Sloan, G.C., Zijlstra, A.A., Mauron, N. \& Houck, J.R., 2012, \mnras, 427, 2588

\bibitem[\protect\citeauthoryear{Lebzelter et al.}{2023}]{Lebzelter2023}
Lebzelter, T., Mowlavi, N., Lecoeur-Taibi, I., Trabucchi, M., Audard, M., Garcia-Lario, P., Gavras, P., Holl, B., Jevardat de Fombelle, G., Nienartowicz, K. et al., 2023, \aap, 674, A15

\bibitem[Lomb(1976)]{1976Ap&SS..39..447L} Lomb, N.~R.\ 1976, \apss, 39, 447. %doi:10.1007/BF00648343

\bibitem[Luna et al.(2013)]{2013A&A...559A...6L} Luna, G.~J.~M., Sokoloski, J.~L., Mukai, K., et al.\ 2013, \aap, 559, A6. %doi:10.1051/0004-6361/201220792


\bibitem[Matsunaga et al.(2017)]{2017MNRAS.469.4949M} Matsunaga, N., Menzies, J.~W., Feast, M.~W., et al.\ 2017, \mnras, 469, 4949. %doi:10.1093/mnras/stx1213

\bibitem[\protect\citeauthoryear{Mauron et al.}{2014}]{Mauron2014}
Mauron, N., Gigoyan, K.S., Berlioz-Arthaud, P. \& Klotz, A., 2014, \aap, 562, A24

\bibitem[\protect\citeauthoryear{Mauron}{2008}]{Mauron2008}
Mauron, N., 2008, \aap, 482, 151

\bibitem[\protect\citeauthoryear{Mauron}{2007}]{Mauron2007}
Mauron, N., Gigoyan, K.S. \& Kendall, T.R., 2007, \aap, 475, 843

\bibitem[Masetti et al.(2011)]{2011A&A...534A..89M} Masetti, N., Munari, U., Henden, A.~A., et al.\ 2011, \aap, 534, A89. %doi:10.1051/0004-6361/201117260

\bibitem[\protect\citeauthoryear{McKeever et al.}{2011}]{McKeever2011}McKeever, J., Lutz, J., Wallerstein, G., Munari, U. \& Siviero, A., 2011, PASP, 123, 1062

\bibitem[Merc et al.(2019a)]{2019RNAAS...3...28M} Merc, J., G{\'a}lis, R., \& Wolf, M.\ 2019a, Research Notes of the American Astronomical Society, 3, 28. %doi:10.3847/2515-5172/ab0429

\bibitem[Merc et al.(2019b)]{2019AN....340..598M} Merc, J., G{\'a}lis, R., \& Wolf, M.\ 2019b, Astronomische Nachrichten, 340, 598.% doi:10.1002/asna.201913662

\bibitem[Merc et al.(2021)]{Merc+2021} 
Merc, J., G{\'a}lis, R., Wolf, M., et al.\ 2021, \mnras, 506, 4151. 
% doi:10.1093/mnras/stab2034

\bibitem[Merc (2025)]{Merc2025} 
Merc, J.\ 2025, Galaxies, 13(3), 49. 
% https://www.mdpi.com/2075-4434/13/3/49

\bibitem[Merrill \& Burwell(1950)]{1950ApJ...112...72M} Merrill, P.~W. \& Burwell, C.~G.\ 1950, \apj, 112, 72. %doi:10.1086/145319

\bibitem[Mikolajewska et al.(1997)]{Mikolajewska+1997} 
Mikolajewska, J., Acker, A., \& Stenholm, B.\ 1997, \aap, 327, 191

\bibitem[Miszalski et al.(2013)]{Miszalski+2013} 
Miszalski, B., Miko{\l}ajewska, J., \& Udalski, A.\ 2013, \mnras, 432, 3186. 
% doi:10.1093/mnras/stt673

\bibitem[\protect\citeauthoryear{Mouhcine \& Lan\c con}{2003}]{ML2003}Mouhcine, M. \& Lan\c con, A., 2003, \mnras, 338, 572

\bibitem[Munari(2019)]{Munari+2019} 
Munari, U.\ 2019, arXiv:1909.01389. % doi:10.48550/arXiv.1909.01389

\bibitem[Munari et al.(1999)]{1999IBVS.4668....1M} Munari, U., Tomov, T., \& Rejkuba, M.\ 1999, Information Bulletin on Variable Stars, 4668, 1

\bibitem[M\"{u}rset et al.(1997)]{Murset1997} M\"{u}rset, U., Wolff, B., \& Jordan, S.\ 1997, \aap, 319, 201

\bibitem[\protect\citeauthoryear{Ortiz \& Guerrero}{2016}]{OG2016} 
Ortiz, R. \& Guerrero, M.A., 2016, \mnras, 461, 3036

\bibitem[\protect\citeauthoryear{Ortiz \& Guerrero}{2021}]{OG2021} 
Ortiz, R. \& Guerrero, M.A., 2021, \apj, 912, 93

\bibitem[Pavlinsky et al.(2022)]{2022A&A...661A..38P} Pavlinsky, M., Sazonov, S., Burenin, R., et al.\ 2022, \aap, 661, A38. 
% doi:10.1051/0004-6361/202141770

\bibitem[Pawlak(2023)]{Pawlak+2023} 
Pawlak, M.\ 2023, \aap, 669, A60. % doi:10.1051/0004-6361/202141782

\bibitem[Pawlak et al.(2024)]{Pawlak+2024} 
Pawlak, M., Trabucchi, M., Eyer, L., et al.\ 2024, \aap, 682, A88. 
% doi:10.1051/0004-6361/202346163

\bibitem[\protect\citeauthoryear{Richter \& Wood}{2001}]{RW2001}Richter, H. \& Wood, P.R., 2001, \aap, 369, 1027

\bibitem[Sahai et al.(2008)]{Sahai+2008} 
Sahai, R., Findeisen, K., Gil de Paz, A., et al.\ 2008, \apj, 689, 1274. 
% doi:10.1086/592559

\bibitem[Sahai et al.(2018)]{Sahai+2018} 
Sahai, R., S{\'a}nchez Contreras, C., Mangan, A.~S., et al.\ 2018, \apj, 860, 105. 
% doi:10.3847/1538-4357/aac3d7

\bibitem[\protect\citeauthoryear{Sahai et al.}{2022}]{Sahai2022}
Sahai, R., Sanz-Forcada, J., Guerrero, M.A., Ortiz, R. \& Contreras, C.S., 2022, Galaxies, 10, 62S

\bibitem[Sanduleak \& Stephenson(1973)]{1973ApJ...185..899S} Sanduleak, N. \& Stephenson, C.~B.\ 1973, \apj, 185, 899. %doi:10.1086/152464

\bibitem[Sanford(1950)]{1950ApJ...111..270S} Sanford, R.~F.\ 1950, \apj, 111, 270. %doi:10.1086/145263

\bibitem[Scargle(1982)]{1982ApJ...263..835S} Scargle, J.~D.\ 1982, \apj, 263, 835. %doi:10.1086/160554

\bibitem[\protect\citeauthoryear{Schlegel et al.}{1998}]{Schlegel1998}
Schlegel, D.J., Finkbeiner, D.P. \& Davis, M., \apj, 500, 525

\bibitem[Seal(1988)]{1988ASSL..145..293S} Seal, P.\ 1988, IAU Colloq. 103: The Symbiotic Phenomenon, 145, 293. %doi:10.1007/978-94-009-2969-2\_61

\bibitem[Shappee et al.(2014)]{2014ApJ...788...48S} Shappee, B.~J., Prieto, J.~L., Grupe, D., et al.\ 2014, \apj, 788, 48. %doi:10.1088/0004-637X/788/1/48

\bibitem[Shetye et al.(2021)]{2021A&A...650A.118S} Shetye, S., Van Eck, S., Jorissen, A., et al.\ 2021, \aap, 650, A118. %doi:10.1051/0004-6361/202040207


\bibitem[Shingles et al.(2021)]{2021TNSAN...7....1S} Shingles, L., Smith, K.~W., Young, D.~R., et al.\ 2021, Transient Name Server AstroNote, 7

\bibitem[Smith et al.(2020)]{2020PASP..132h5002S} Smith, K.~W., Smartt, S.~J., Young, D.~R., et al.\ 2020, \pasp, 132, 085002. %doi:10.1088/1538-3873/ab936e

\bibitem[Sobotka et al.(2003)]{2003ASPC..303..171S} Sobotka, P., Pejcha, O., {\v{S}}melcer, L., et al.\ 2003, Symbiotic Stars Probing Stellar Evolution, 303, 171. %doi:10.48550/arXiv.astro-ph/0210585

\bibitem[Soszy{\'n}ski et al.(2021)]{Soszynski+2021} 
Soszy{\'n}ski, I., Olechowska, A., Ratajczak, M., et al.\ 2021, \apjl, 911, L22. 
% doi:10.3847/2041-8213/abf3c9

\bibitem[Tonry et al.(2018)]{2018PASP..130f4505T} Tonry, J.~L., Denneau, L., Heinze, A.~N., et al.\ 2018, \pasp, 130, 064505. %doi:10.1088/1538-3873/aabadf

\bibitem[van Belle et al.(1997)]{1997AJ....114.2150V} van Belle, G.~T., Dyck, H.~M., Thompson, R.~R., et al.\ 1997, \aj, 114, 2150. %doi:10.1086/118635


\bibitem[van Belle et al.(2013)]{2013ApJ...775...45V} van Belle, G.~T., Paladini, C., Aringer, B., et al.\ 2013, \apj, 775, 1, 45. %doi:10.1088/0004-637X/775/1/45


\bibitem[Webb et al.(2020)]{Webb+2020} 
Webb, N.~A., Coriat, M., Traulsen, I., et al.\ 2020, \aap, 641, A136. 
% doi:10.1051/0004-6361/201937353

\bibitem[Wetuski et al.(2021)]{2021MNRAS.506.5619W} Wetuski, J., Hynes, R.~I., Maccarone, T.~J., et al.\ 2021, \mnras, 506, 5619. 
% doi:10.1093/mnras/stab2139

\bibitem[Whitelock(1987)]{1987PASP...99..573W} Whitelock, P.~A.\ 1987, \pasp, 99, 573. %doi:10.1086/132019

\bibitem[\protect\citeauthoryear{Whitelock et al.}{2006}]{Whitelock2006}Whitelock, P.A., Feast, M.W., Marang, F. \& Groenewegen, M.A.T., 2006, \mnras, 369, 751

\bibitem[Wilms et al.(2000)]{Wilms2000} Wilms, J., Allen, A., \& McCray, R.\ 2000, \apj, 542, 914.

\bibitem[Wood et al.(2004)]{Wood+2004} 
Wood, P.~R., Olivier, E.~A., \& Kawaler, S.~D.\ 2004, \apj, 604, 800. 
% doi:10.1086/382123

\bibitem[\protect\citeauthoryear{Wo\'zniak et al.}{2004a}]{Wozniak2004a}Wo\'zniak, P.R. et al., 2004, \aj, 127, 2436

\bibitem[\protect\citeauthoryear{Wo\'zniak et al.}{2004b}]{Wozniak2004b}Wo\'zniak, P.R., Williams, S.J., Vestrand, W.T. \& Gupta, V., 2004, \aj, 128, 2965

\bibitem[\protect\citeauthoryear{Wright et al.}{2010}]{Wright2010}
Wright, E.L. et al., 2010, \aj, 140, 1868

\bibitem[Zhao et al.(2012)]{2012RAA....12..723Z} Zhao, G., Zhao, Y.-H., Chu, Y.-Q., et al.\ 2012, Research in Astronomy and Astrophysics, 12, 723. %doi:10.1088/1674-4527/12/7/002


\end{thebibliography}
%
% - join the .bib files when you upload your source files
%-------------------------------------------------------------------

\end{document}